\documentclass[]{interact}

\usepackage{verbatim}
\usepackage{epstopdf}
\usepackage{subfigure}

\usepackage{pdfpages}

\theoremstyle{plain}

\theoremstyle{definition}

\theoremstyle{remark}

\begin{document}

\articletype{article}

\title{TAPS Responsibility Matrix: A tool for responsible data science by design}

\author{
\name{Visara Urovi, \textsuperscript{a}\thanks{CONTACT Visara Urovi. Email: v.urovi@maastrichtuniversity.nl}, Remzi Celebi \textsuperscript{a},Chang Sun\textsuperscript{a}, Linda Rieswijk\textsuperscript{a}, Michael Erard\textsuperscript{a}, Arif Yilmaz\textsuperscript{a}, Kody Moodley\textsuperscript{a}, Parveen Kumar\textsuperscript{a} and Michel Dumontier\textsuperscript{a}}
\affil{\textsuperscript{a}
Institute of Data Science, Maastricht University, P.O. Box 616, 6200 MD, Maastricht, The Netherlands;
}
}

\maketitle

	\begin{abstract}
Data science is an interdisciplinary research area where scientists are typically working with data coming from different fields. When using and analyzing data, the scientists implicitly agree to follow standards, procedures, and rules set in these fields. However, guidance on the responsibilities of the data scientists and the other involved actors in a data science project is typically missing.
While literature shows that novel frameworks and tools are being proposed in support of open-science, data reuse, and research data management, there are currently no frameworks that can fully express responsibilities of a data science project. In this paper, we describe the Transparency, Accountability, Privacy, and Societal Responsibility Matrix (TAPS-RM) as framework to explore social, legal, and ethical aspects of data science projects. TAPS-RM acts as a tool to provide users with a holistic view of their project beyond key outcomes and clarifies the responsibilities of actors. We map the developed model of TAPS-RM with well-known initiatives for open data (such as FACT, FAIR and Datasheets for datasets).  We conclude that TAPS-RM is a tool to reflect on responsibilities at a data science project level  and can be used to advance responsible data science by design. 
	\end{abstract}
	
\begin{keywords}
Responsible Data Science , Responsibility framework, Transparency, Privacy/Confidentiality, Accountability, Societal Values
\end{keywords}

	\maketitle
	
	\section{Introduction}
	
	Data Science is an emerging interdisciplinary domain whose responsibility agenda has largely been borrowed or reconfigured from other domains. Most data scientists work in teams where typically data come from different fields (such as business, finance, healthcare).
	To access and work with the data, data scientists are involved in 'data prospecting' \cite{slotaProspectingDataSciences2020a}, a process requiring the data scientists to explicitly and implicitly agree to abide by the responsibility agenda of these domains. For example, working with biomedical, clinical, and other health-related data requires protecting the anonymity and confidentiality of individual patients by means of an informed consent and other principles and norms \cite{kalkmanResponsibleDataSharing2019a}. In general, working with data and data analysis also requires engaging and defending the necessary trade-offs between individuals\textquoteright{} privacy and public's interest in research results. Even if some concerns about data sharing were raised in the past  \cite{longoDataSharing2016}, 
	arguably, the benefits of open science and data sharing have been widely recognized in various research domains \cite{byrdResponsiblePracticalGenomic2020, moorthyEmergingPublicHealth2020}. In data science projects however, once the data is available, some questions remain hard to answer such as: 
	How data and algorithms will be made transparent?
	What are the social and ethical implications of the outcomes of the project? Is there a procedure for dealing with unintended outcomes? 
	We argue that these questions should be answered and documented during the project life cycle so that the work and the results persist beyond the timeline of the project itself. 
	
	Wil van der Aalst was one of the first to write about Responsible Data Science \cite{vanderaalstGreenDataScience2016, van2018responsible}. 
	He refers to data being the new oil (now a familiar metaphor) and the use of data science methods to create new forms of energy. He mentions that in order to resolve these \textquotedblleft pollution\textquotedblright{} problems caused by systematic discrimination based on data, invasions of privacy, non-transparent life-changing decisions, and inaccurate conclusions, data scientists should develop technological solutions that ensure fairness, confidentiality, accuracy, and transparency, today known as the FACT principles.  Since then, more attention has been paid to ethics, fairness, and equity in artificial intelligence, machine learning, and big data. Machine learning approaches are generally developed as black boxes, and the source code for these models is rarely inspected in terms of fairness, accuracy, accountability, and transparency \cite{piano2020ethical}. Popular understandings of data science and data analytics have been negatively influenced by revelations of companies collecting and using big data illegitimately. Further, revelations of algorithmic bias have shown the need for data scientists to foresee in advance the possibility for misuse of the technologies they build. 
	In this work we view responsibility as more than simply complying with legal regulations and ethical requirements; it also means that it is neither an incidental characteristic of a data science project nor an accidental byproduct. Instead, we refer to responsibility 'by design' as a reflection on two important components of data science projects: 1) foresight and 2) infrastructural support. 
	\begin{itemize}
	\item {Foresight:} Achieving responsibility involves a level of intentionality in the planning and execution of a data science project. What are the potential pitfalls? Are the planned benefits of a project? Achieving this requires a level of ethical self-reflection by researchers and the involved organisations about their preparedness to make certain types of decision, as well as honest assessments of their ability to execute. This self-reflection also requires self-assessments about the completeness of one\textquoteright s understanding about the impact of technology. 
	\item {Infrastructural support} \textquotedblleft By design\textquotedblright{} also refers to the
	way responsibility is built into all aspects of a data science enterprise, from how people are trained to how research projects are budgeted  and supported by infrastructures and organizations. Achieving responsibility should not be placed solely on individuals, but should be woven into a range of everyday practices.
	\end{itemize}
	To achieve responsibility by design, we propose the TAPS Responsibility Matrix (TAPS-RM), a framework for defining responsibility in data science projects. We identify four components of responsibility, namely transparency, privacy and confidentiality, accountability and societal values, and their scopes which are used as a basis for defining responsibility. The responsibility matrix proposed here is viewed as a generic framework for thinking about and identifying responsibility in data science organisations/projects, rather than prescribing specific normative constraints about what responsibility should look like in these organisations. A similar concept for project management and coordination was described by RACI matrix \cite{haughey2017raci}, which outlines the roles and responsibilities of involved actors. Our matrix, in contrast, focuses on broader social and ethical responsibilities within a project (i.e., ELSI \cite{FISHER2005321} aspects). Some of the responsibility aspects that we cover have also been described in other frameworks such as FAIR \cite{wilkinsonFAIRGuidingPrinciples2016}, Datasheets for Datasets \cite{gebruDatasheetsDatasets2021} or Model Cards \cite{ModelCards2018}, here we provide the definition of responsibility as explicit components relating to actors, objects, processes and impacts. The contribution of the work is twofold: a generic framework to be used for defining the responsibilities in a data science project and a framework where other established ethics and responsibility frameworks can be aligned, to provide a structured view and clarify the scope of the responsibilities in a project. We compare the responsibility matrix with FACT, FAIR and Datasheets for datasets as alternative frameworks for addressing responsibility in data driven projects. We show that the proposed matrix is a generic and applicable tool that can be used in Data Science projects.
	
	The remainder of this work is organised as follows: In Section \ref{sec:related} we provide an overview of the state of the art works on open data-sharing and data-driven development approaches. In Section \ref{sec:matrix} we provide an overview of the components and dimensions of the Matrix, our definitions and how it all works together to define a Responsibility by Design approach. We exemplify our approach with a case study. In section \ref{sec:evaluation} we follow up with a comparison of the matrix with well-known approaches to responsible data science and in section \ref{sec:discussion} and proceed with a discussion of the approach. Finally, section \ref{sec:conclusion} concludes with a summary of the main findings and future works.
	
	\section{Related Work}
	\label{sec:related}
	
	To date, several existing frameworks and initiatives attempt to address responsibility issues that arise within data science. Since 2014, the social science FAT{*} community (organized by the Association for Computing Machinery (ACM)) has introduced the FACT principles in order to ensure fairness, accuracy, confidentiality and transparency. FACT principles signify efforts to protect human subject's privacy and confidentiality in line with legal data protection compliance
	and reduce bias in models and data and provide more accurate analysis \cite{vanderaalstResponsibleDataScience2017}. Taylor and Purtova \cite{taylorWhatResponsibleSustainable2019} argue the data should be a commons, therefore theory of a common-pool-resource, establishing conditions under which the shared resource is used sustainably,
	is relevant to enable responsible data science. In 2016, the FAIR data principles were introduced as a way to improve the machine-actionability (i.e., the capacity of computational systems
	to find (Findability), access (Accessibility), interoperate (Interoperable), and reuse (Reusability) data without or with minimal human intervention). In 2018, the CARE Principles for Indigenous Data Governance \cite{carroll2020care}, which stands for Collective Benefit, Authority to Control, Responsibility, and Ethics, were added to complement the FAIR principles by the Global Indigenous Data Alliance. The CARE Principles for Indigenous Data Governance are people and purpose-oriented, reflecting the crucial role of data in advancing Indigenous innovation and self-determination. These principles complement the existing FAIR principles encouraging open and other data movements to consider both people and purpose in their advocacy and pursuits. 
	
	Large companies such as Microsoft and Google have also proposed their own solutions to tackle all kinds of responsibility issues. Datasheets for Datasets \cite{gebruDatasheetsDatasets2021} were introduced by Microsoft \cite{Microsoft2018} to improve documenting of datasets used for training and evaluating machine learning models. The aim of the work is to increase dataset transparency and facilitate better communication between dataset creators and dataset consumers (e.g., those using datasets to train machine learning models). They provide a set of questions that should be answered. Google launched a similar initiative specifically for machine learning algorithms in 2018, which introduced the concept of \textquotedblleft Model Cards\textquotedblright{} \cite{ModelCards2018}. This effort aims to help organize the essential facts of machine learning models in a structured way. 
	In December 2020, Rolls-Royce also introduced their Aletheia FrameworkTM \cite{RollsRoyce2020} which
	is a practical toolkit that helps organisations to consider the impacts on people of using artificial intelligence prior to deciding whether to proceed. It looks at 32 aspects of social impact, governance
	and trust, and transparency and requires executives and boards to provide evidence that these have been rigorously considered \cite{RollsRoyce2020}. Yet, these three are representative of efforts to promote responsibility in data science: each was developed by experienced practitioners with extensive backgrounds in responsibility; each is mainly prescriptive but is not tied to enforceable penalties (except in the case of compliance with GDPR); none is reflective of all the variables that influence data science projects in academia and business. Finally, these frameworks are neither competitors nor collaborators; they are more like the proverbial blind men and the elephant, each offering perspectives that are usually incommensurable with each other. These frameworks and discussions informed our thoughts, and out of this exploration we designed a tool that is both analytic and productive, in the sense that it can be used to analyze and diagnose existing projects as well as to help make nascent projects more responsible: we call this tool TAPS responsibility matrix. 
	
	\section{The Responsibility Matrix }
	\label{sec:matrix}
	
	In this section, we introduce the approach and the dimensions of the responsibility matrix, the four scopes (actors, objects, processes and impacts) and the four components (transparency, accountability, privacy/confidentiality and societal values). Within the two-dimensional space of the matrix arise numerous spaces for asking questions and making recommendations to make data science projects responsible. It clarifies discussions about responsibility and raises topics which often stay
	implicit. It also stimulates thinking and accountability along dimensions that data scientists may be prone to overlook. 
	
	\subsection{Matrix development approach}
	
	We started our Matrix development with a small working group (10 people) which identified the main dimensions of responsible approaches to data science which uniformly \textquotedblleft tile the space\textquotedblright{} of responsible data science by design. Initially, we chose the categories that were stable for a range of principles and practices that exist and/or are emerging in data science. Then we developed and tested the descriptions of the components in brainstorm meetings with data scientists, modifying some language as a result. For example, \textquotedblleft accountability\textquotedblright{} was first called \textquotedblleft governance,\textquotedblright{}
	and  \textquotedblleft confidentiality\textquotedblright{} was later added to the \textquotedblleft privacy\textquotedblright{} label. Once we agreed on four categories, we conducted targeted interviews with 6 external data scientists to test the thematic coherence of the identified dimensions. 
	
	\textbf{Thematic coherence.} The generated definitions of the components were tested in semi-structured interviews (n=6) with data scientists working in academic and business domains. Based on their experience with specific projects, they provided feedback on the sufficiency of our descriptions of the components. Specifically, we asked interviewees to describe instances of projects and checked that all of these instances could fit into the identified components. All of the participants found the descriptions to be relevant and thematically coherent, but they provided feedback about shortcomings. For instance, we learned the importance of addressing liability in the \textquotedblleft accountability\textquotedblright{} component. 
	We did not explicitly test the scope definitions (agents, objects, processes impacts) with users. The interviews gave us confidence that the four components capture all important dimensions of responsible data science by design. 
	
	\textbf{Usability.} The resulting two-dimensional matrix (components
	x scopes, producing 16 cells at the intersection) were disseminated
	informally to the Institute of Data Science Team at Maastricht University.
	Twenty participants were asked to attempt to map one of their current
	projects onto the matrix. From this, we learned that the data scientists
	1) need prompting questions in order to fill in the cells of the matrix
	with appropriate material and 2) they need directions in which components, scopes, and individual cells to address first. Based on the feedback and the definitions of each cell, we added several questions to guide the process of filling in the matrix. 
	
	\textbf{Applicability.} We evaluated the matrix by mapping existing frameworks for ethics and responsibility in data science, computer science, and artificial intelligence into the specific cells. The frameworks we selected were the FAIR guidelines (gofair.org), the FACT framework \cite{vanderaalstResponsibleDataScience2017}, and Datasheets for Datasets \cite{gebruDatasheetsDatasets2021}. We chose to evaluate the matrix with these three frameworks because they ranged from concisely formalized (FAIR) to expansive (Datasheets for Datasets), and because their creators ranged from international consortia (FAIR, Datasheets for Datasets) to single authors (FACT). 
	More specifically, the FAIR guidelines were selected because they provide increasingly important guidance on an international level to how researchers produce digital research objects. Experiences of our group in their dissemination and implementation of FAIR also inform the formulation of the concept of the responsibility matrix.
	The FACT guidelines were chosen because they have been prominently
	circulated in research circles. The \textquotedblleft Datasheets for
	Datasets\textquotedblright{} guidelines were an intriguingly comprehensive approach that mirrored the responsibility matrix in its use of open questions that researchers would ask themselves. 
	
	\subsection{The four scopes of responsibility}
	
	We identified four scopes over which the components of a responsibility
	agenda operate: actors/agents, objects, processes, and impacts. 
	
	\textbf{Actors/agents:} are the people and/or legal entities who are involved in data science projects, such
	as researchers, end users, data subjects, annotators, Mechanical Turk
	contractors, organizations, etc \footnote{ We discussed the possibility of including artificial intelligence as an actor as well. We decided against it at this time because they are not considered legal persons. When their status changes, we will revisit this definition.}. They bear responsibilities in the project (identified by their role in it). 
	
	\textbf{Objects:} are data products or digital resources, such as
	prediction models, databases, knowledge graphs, etc. Essentially,
	an object is anything that would be considered a deliverable for a
	client or funding agency. 
	
	\textbf{Processes:} are the set of plans, activities, or steps taken
	by the actors to develop and test the objects. For example, the canonical
	components of the data science lifecycle, experiments, trials, standard
	operating procedures.
	
	\textbf{Impacts:} are the intended and unintended outcomes of a data
	science project. They provide a way of reflecting on the suitability of the performed research question. For example, what change does it make in the world? How does it
	impact people\textquoteright s lives? How does it impact innovation and economy?
	
	\subsection{The Four Components of Responsibility}
	
	The four components are transparency, privacy/confidentiality, accountability, and societal values. Bellow we describe these components and the role they play in a responsibility agenda. Later we develop specific recommendations for each component.
	
	\subsubsection{Transparency} 

	Transparency, in general, plays an important role not only in the reproducibility of a study, but also in avoiding unwanted/unintended consequences and misinterpretations. Transparency allows other researchers to check the validity of scientific methods, to ensure that desired results can be achieved and to prevent misinterpretation of the results. Lepri et al. 2018 emphasizes the
	understandability of machine models and defines transparency as a
	mechanism facilitating algorithmic accountability \cite{lepriFairTransparentAccountable2018}. There are opinions
	suggesting that publication of datasets and other open-access metadata
	can bring gains in transparency, accountability and fairness \cite{lathropOpenGovernmentCollaboration2010}. Governments began to use open standards and publish their data for the ideal of being fully transparent. At the same time, the limits of transparency have been unclear and the subject of debate \cite{kemperTransparentWhomNo2019a} Annany and Crawford suggest that transparency cannot be solely a property of an algorithmic model, but should instead be considered in the context of socio-technical interactions between
	algorithms and people \cite{anannySeeingKnowingLimitations2018}. 
	Our definition of transparency for data science reflects this context. We define transparency in four scopes:
	\begin{itemize}
	    \item \textbf{Actors:} This refers to identifying the roles, the contributions and the interest of the involved actors (i.e. principal investigator, developers, end users, organisations). 
	    \item \textbf{Objects:} This means describing the detailed provenance of the objects (i.e metadata, license, quality measures, bias in data) are used and/or generated in the project.
	    \item \textbf{Processes: }This means describing the processes (i.e. methodology, interaction protocols, pre-registering or business processes) employed in the project.
	    \item \textbf{Impacts:} This refers to identifying and describing the short, and long term impacts (both negative and positive) of the project (i.e. key exploitable results, recognisable possible negative and positive downstream impacts).
	\end{itemize}
	In addition, intended uses, other possible uses, possible biases and
	limitations of a data science project should be clearly documented
	and communicated to different actors using understandable language
	and terminology (i.e see \cite{celisInterventionsRankingPresence2020a}.

\subsubsection{Privacy/Confidentiality}

Modern data societies are evolving the sense that certain uses of data might challenge the autonomy of data subjects. Privacy and Confidentiality of the data are the two key concepts for autonomy. 

\textbf{Privacy} has been defined from various points of view, such as the law, organisation, and philosophy. In this context we define privacy as the right and the freedom of a person to determine which personal information about himself/herself may be shared and accessed by others~\cite{jacksonReviewPrivacyFreedom1968,bennPrivacyFreedomRespect1971,harrissUpdateEthicalStandards2011,derlega1977privacy}.
	
\textbf{Confidentiality} refers to personal and non-personal information shared with a select group or for an explicit purpose that generally cannot be divulged to third parties without the expressed consent of the individual data subject. This consent is granted with the expectation that it will not be revealed to others or used for non-compatible purposes. While confidentiality is an ethical duty, privacy is a right rooted in the common law. For example, the General Data Protection Regulation 2016/679 \cite{regulation2016regulation} is a regulation in the EU law on data protection and privacy in the European Union and the European Economic Area. GDPR is a mechanism for satisfying these ideas about the autonomy of the data subject. 
	
	We define privacy and confidentiality in terms of actors, objects,
	process and impacts: 
	\begin{itemize}
	    \item \textbf{Actors: }Identifies the privacy and confidentiality agreements between the different actors of the project (i.e researchers, end users, data subjects, organisations). For example, do data collectors inform human data subjects about the collection, use and storage of their data and to obtain their consent (i.e. an informed consent agreement)? For human data processors or controllers what confidentiality agreements have been signed in the project (i.e. Non-Disclosure Agreement). 
	    \item \textbf{Objects: }Identifies sensitive and/or confidential data. For example, is there any data or knowledge that should not be used or disclosed? 
	    \item \textbf{Processes:} Explicits the processes of the project that ensure data privacy and confidentiality. For example, what process is followed to update or delete records on the request of data subjects? Which security protocols or privacy-preserving methods are employed? 
	    \item \textbf{Impacts: }Identifies if and how the project impacts the privacy and confidentiality of the data beyond the project lifecycle. For the identified impacts, which measures are taken to ensure privacy and confidentiality. For example, how the project outcomes and the supporting data will be shared with third parties while preserving privacy of data subjects?
	\end{itemize}
	
\subsubsection{Accountability}
		
Accountability can be defined as those considerations within data science projects associated with crediting or charging someone for an
action with a recognized responsibility, that is associated with a role or a set of functional or moral obligations. 
Accountability also pertains to how a particular responsibility agenda is implemented and empowered within the structure of an organisation. An organisation shows its commitments to supporting responsible data science in its internal ethical stances and disciplinary processes, its organisation of teams, and budgetary priorities. Unpacking this definition in relation to internal ethical stances of organisations, topics for consideration include governance procedures and criteria for establishing clear social, legal and ethical codes to which each member in a project must abide, for example ethical guidelines in healthcare data processing \cite{bentonEthicalResearchProtocols2017}. Disciplinary processes concern the corrective and reactive measures an organisation takes, and their procedures for dealing with individuals involved in deliberate and persistent malpractice. Related to this matter are the criteria and processes for the assembling of teams to carry out specific data science activities for an organisation and, more specifically, the definition and mapping of roles and data responsibilities for all individuals and groups in the data science pipeline. 
Preventative (as opposed to purely reactive) measures for accidental data malpractice also fall under the accountability dimension. Training and educational programs can be established by organisations in order to raise awareness and to provide knowledge to help staff prevent accidental malpractice in the particular part of a data science project they lead or conduct. These programs can also educate staff on what potential social, legal and ethical pitfalls each process in the data science pipeline is prone to. 
Ultimately, accountability depends on the gravitas with which a society views transparency, privacy, confidentiality, and the preservation of its values in data-related projects. It is also based on the extent to which societies want data to be useful, what kinds of data it considers useful, and for what purposes it considers them useful. The sense of accountability we define here is ascribed to all levels of resolution of an organisation, in this way similar to Pearson's notion of accountability \cite{pearsonStrongAccountabilityIts2017}.
	
For each scope, we define accountability as follows:
	\begin{itemize}
	    \item \textbf{Actors:} This defines the accountability frameworks (i.e contracts and agreements) used to assess or prescribe the responsibilities of each actor in relation to the project and its outcomes. For example, what is the principal investigator accountable for and what are the obligations of the leading organisation. 
	    \item \textbf{Objects:} This identifies the quality (e.g. being bias-free, anomalous or interoperable), availability and terms of use of the objects. For example: Is data and its provenance available; Is there a description of the biases in the data; What is the license attached to the data? Who is accountable for the quality, availability and terms of use of the objects? 
	    \item \textbf{Processes:} This defines the processes that are in place for monitoring and ensuring an accurate, ethical and lawful functioning of the project (i.e. ethical approval) 
	    \item \textbf{Impacts:} A project can have various types of short, middle and long term outcomes. These outcomes can bring about downstream impacts and consequences that are either positive or negative. This identifies the procedures for dealing with the ramifications of the outcomes (i.e intellectual property management or responsibilities of all actors in relation to negative outcomes)
	\end{itemize}
	
\subsubsection{Societal Values}
	
The societal values dimension of our responsibility matrix concerns the fact that societies differ in which norms and values they prioritize, and societies often hold inconsistent or paradoxical norms and values. These norms and values can also change, often in unpredictable ways. Therefore, projects must also consider the operative values of the particular society in which it is embedded and those that it interacts with.
These ethical norms and societal values can involve the preservation of human life, the importance placed on innovation and technological progress, the cohesion of communities, the improvement of health, the sovereignty of individuals and their freedom to choose, and so forth. Therefore, exactly how societal values are taken into account in the execution of data science, and how these values affect, qualify or modify processes and goals in data science activity varies across societies. As far as data science goes, projects are often undertaken to support and respect those values. The pursuit of these norms and values are often used to motivate data science projects, and they are used to justify trade-offs between interests that may be at odds with each other. Examples of how societal values are instantiated in data science projects are the German ethics code for autonomous and connected driving \cite{luetgeGermanEthicsCode2017} and the moral machine experiment \cite{awadMoralMachineExperiment2018a} which tries to curate and understand data about the consensus of a particular society with regards to specific moral or ethical dilemmas. 
	We define societal values in four scopes; 
	\begin{itemize}
	    \item 	\textbf{Actors:} This refers to identifying the operative societal values and norms of the actors, and how they are addressed and/or protected (i.e elicit values of the work, identify the social context that applies to the work. 
	    \item \textbf{Objects:} This means explicitly applying societal values and norms to the objects (i.e. avoiding specific biases that may arise from the produced algorithms). 
	    \item \textbf{Processes: }Identifying the processes that are in place to address societal values (e.g. how are possible biases addressed?); \item \textbf{Impacts: }This refers to identifying how outcomes affect or interact with the identified societal values and norms (i.e., who could stand to benefit and who would lose from the outcomes of the work?). And are there any mitigation strategies/laws/rules/communications in place to ensure that the outcomes are responsibly used and exploited for the benefit of society?).
	\end{itemize}

	\subsection{The Responsibility Matrix }
	
The result of applying the components of responsibility across the scopes of responsibility is a two-dimensional matrix of spaces that we call a 'responsibility matrix'. Fig. \ref{TAPSMatrix} shows the responsibility matrix and defines each cell of the 2-dimensional space:
    
    \begin{figure}
	\centering
	\includegraphics[width=\linewidth]{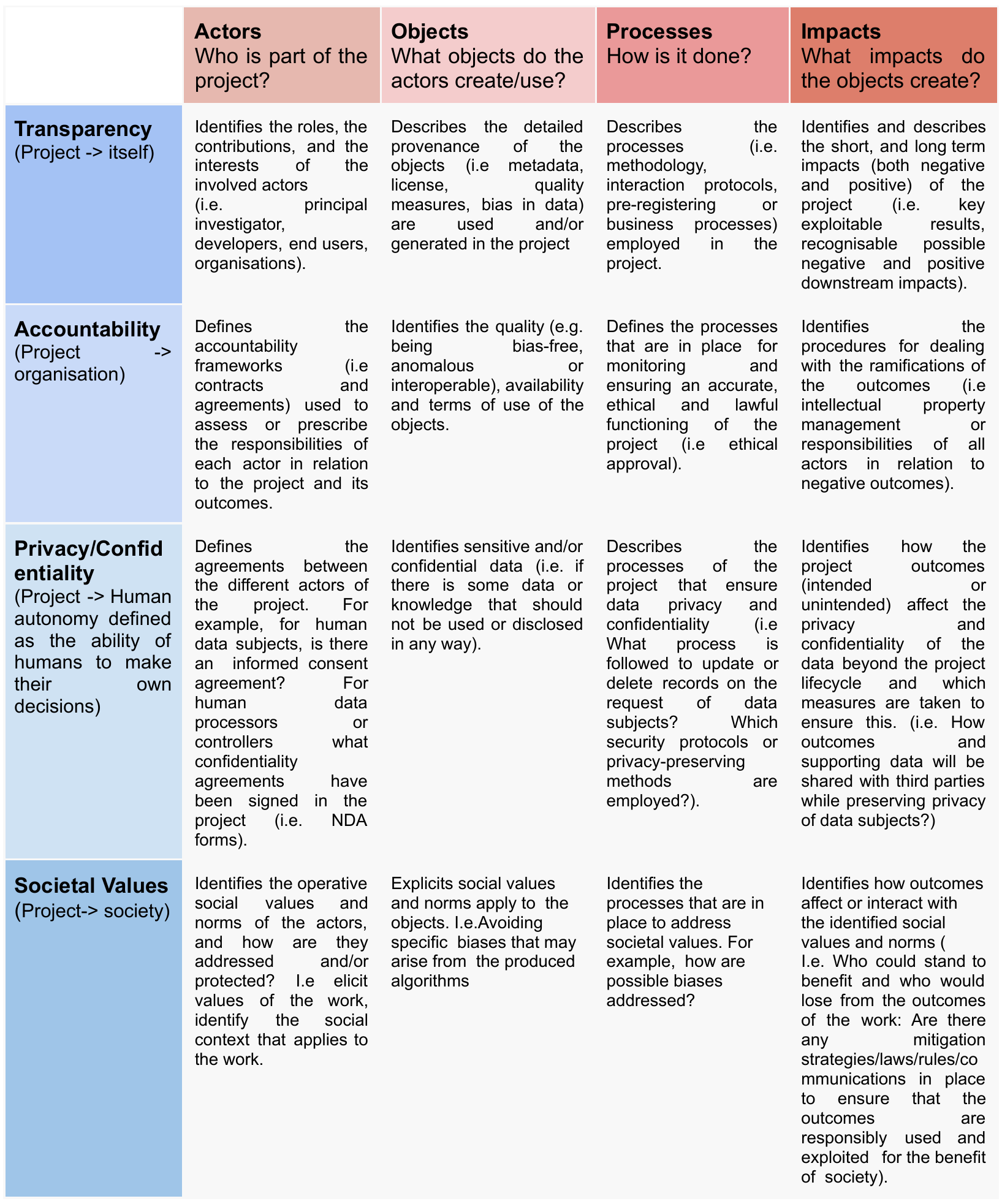}
    \caption{\bf The responsibility matrix as a result of applying the components of responsibility across the scopes of responsibility}
	\label{TAPSMatrix}
    \end{figure}

	
	A set of questions are defined for providing information for each
	cell of the Matrix. A full description of the questions within the
	Responsibility Matrix is attached in Annex A. 
	
	\subsection{Case Study Example}
	
	To show-case the full extent of the TAPS-RM  we describe a  data science project involving two organisations. This hypothetical project consists of developing a predictive model for the prediction of heart failure mortality.  We chose this example because it enables us to (i) use this example to train staff in our organisation to use the matrix and (iii) identify if there are differences in two highly related organisations. This is due to the fact that the medical collaborators are also employed at the faculty of health and medicine of the same university, thus enabling a reflection over the guidelines across departments. The responsibility matrix of multiple organisation projects is very similar and still requires organisations to identify and shape responsibilities into the project. 
    The proposed project is a collaboration between the Data Science Department (IDS) and the Mastricht Medical Center Hospital (MUMC).The goal is to analyse a retrospective cohort of Heart Failure Patient Data. The data are collected from primary care. The aim is to build and deploy a predictive model for heart failure mortality risks. The model will help patient stratification and will allocate more attention to high risk patients. Deep Learning Techniques will be used to analyse textual doctors notes  and structured data (such as gender and comorbidities) on the medical records of patients. A later randomised clinical trial will be performed to test the model with 100 patients.
    
    \begin{figure}
    \centering
	\includegraphics[width=\linewidth]{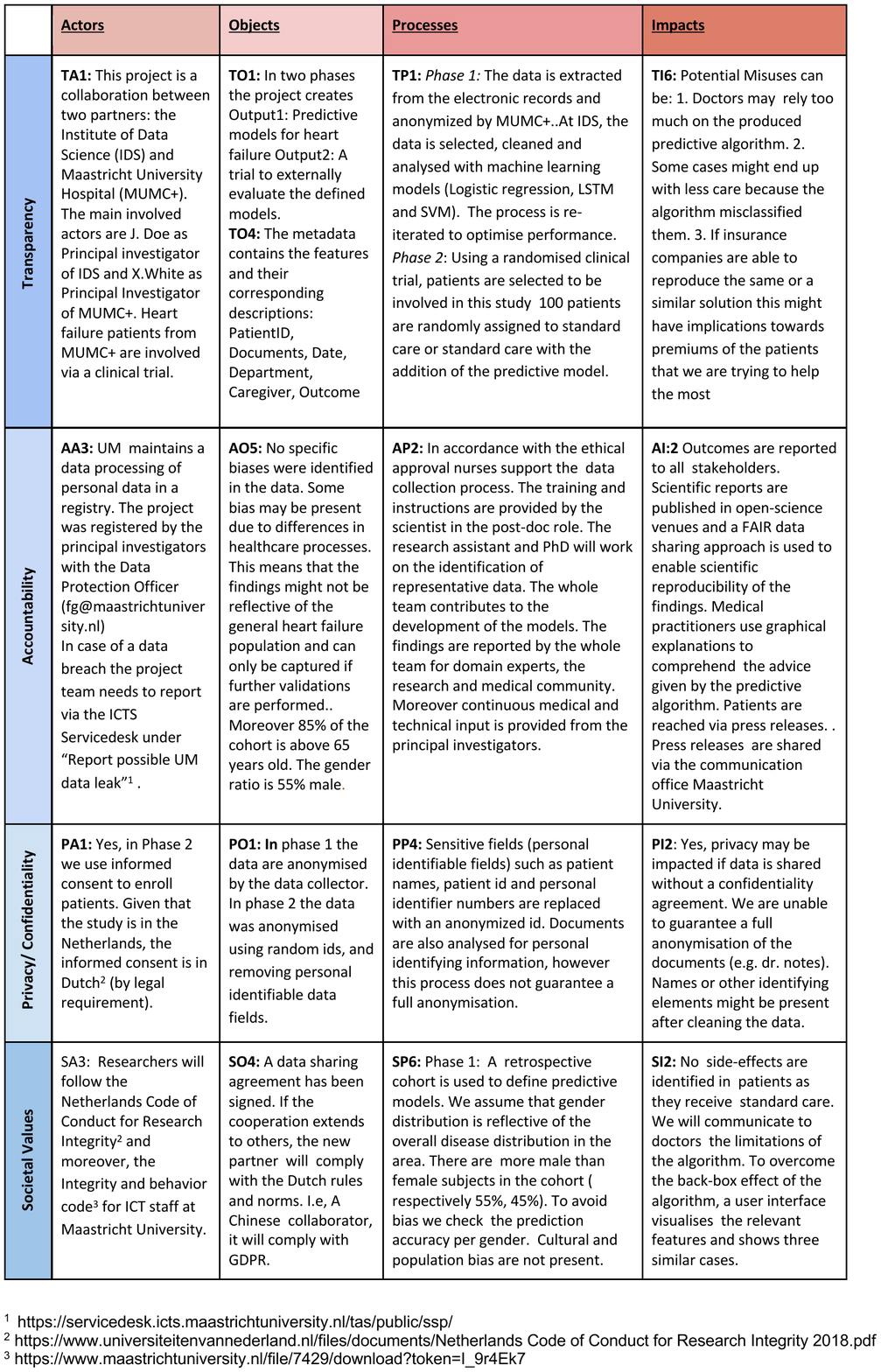}
    \caption{\bf Applying the responsibility matrix to a research project as a case study example.}
    \footnotesize\emph{*} \cite{van2018using,sun2018analyzing,wouters2021putting},
    \footnotesize\emph{**} \cite{wouters2021putting}
	\label{TAPSCaseStudyTable}
    \end{figure}
	
	To exemplify the matrix, in Fig. \ref{TAPSCaseStudyTable}, we show some of the key information of the defined in for this project in the responsibility matrix. The figure shows answers to some of the main questions for each dimension. The answered questions are indicated in each cell and refer to the question in Appendix A. The full responsibility matrix was filled using the web-page \cite{jansenFAIRPrinciples1}, a preliminary user interface developed for filling in the responsibility matrices for different projects. 
	
	\section{Evaluation and Results}
	\label{sec:evaluation}
	We compared the responsibility matrix with FAIR\cite{wilkinsonFAIRGuidingPrinciples2016}, FACT \cite{vanderaalstGreenDataScience2016,vanderaalstResponsibleDataScience2017} and Data-Sheets for Datasets \cite{gebruDatasheetsDatasets2021} which are regarded as three well-known frameworks working towards open data science practices. Although some frameworks are prescriptive and others are designed to be descriptive, they  contain strong elements of responsibility in them or attribute to raising awareness about responsibility.
	The principles, guidelines, and recommendations from the three frameworks are located in the matrix. The matrix format shows where these frameworks overlapped with each other and where gaps exist. The mapping for each framework was made independently by teams of two, if the mapping did not match then it was discussed in the group.
	
	Two researchers independently mapped the chosen guidelines into the matrix. Then, they compared the answers and identified discrepancies. The goal was not to resolve these discrepancies in judgment but rather to identify them and note them as places where either 1) the responsibility matrix was ambiguous or 2) the specific guidelines were ambiguous. 
	
	\subsection{FAIR Principles}
	
	The FAIR data principles recommend a set of guidelines on how to share research data in such a way that it can be findable, accessible, interoperable, and reusable \cite{wilkinsonFAIRGuidingPrinciples2016}. These principles
	have now been adopted by many European research institutions. Making the research data FAIR is the first and the most critical step towards transparency of data-dependent projects. The core set of principles
	can be applied to any digital objects including workflows, software, and research output. 
	\begin{itemize}
    	\item The FAIR principles require \underline{clarity and transparency around the conditions} \underline{governing access and reuse}.
	
		\item FAIR data are required to have a \underline{clear, preferably machine readable, license. }
		\item \underline{The transparent but controlled accessibility of data and services} allows the participation of a broad range of sectors.
	\end{itemize}
	There are 15 FAIR sub-principles \cite{wilkinsonFAIRGuidingPrinciples2016,jansenFAIRPrinciples1}, distributed over four main themes: Findability, Accessibility, Interoperability, and Reusability (Appendix \ref{appendixB}).
	We mapped these 15 sub-principles to the appropriate cells in the matrix. Most of the FAIR sub-principles can appropriately be placed into cells at the intersections of Transparency/Object and Accountability/Object as these principles aim to increase share and reuse of digital objects.
	We map sub-principles for Findability (F1, F2, F3) to the Transparency/Object cell. Additionally, F4 can be mapped to Accountability/Object to the degree that someone/an organisation is tasked with registering and indexing data. Accessibility principles (A1, A1.1, A1.2, A2) ensure that users have access to necessary (metadata) data, emphasizing the need to know how to access them, including authentication and authorization.   A1, A1.1, A1.2 and A2 correspond to the Transparency/Process cell as these principles describe the mechanism to make the data available and accessible to users. Interoperability
	principles (I1, I2 and I3) in FAIR describe the need for data to work
	with applications or workflows for analysis, storage, and processing.
	These principles are in line with our Accountability/Object cell.
	The ultimate goal of FAIR is to optimize data reuse, which is indicated
	in Reusability principles. To achieve this, metadata and data should
	be well-described so that they can be replicated and/or combined in
	different settings. These principles (R1, R1.1, R1.2 and R1.3) are
	in line with the Societal Values/Object cell, where data can be an asset
	if the data is reusable by others. In Fig. \ref{fig:1}, we show in blue how FAIR is mapped to the matrix. While most of the mappings were clearly placed, some cells contained multiple sub principles while some contained
	none. We can conclude that the FAIR principles are focused on the Transparency, Accountability, and Societal Values components of a responsibility agenda, and the Process and Object scopes. Yet there are no sub-guidelines related to the Actors or Outcomes scopes, nor to the Privacy/Confidentiality component. Interestingly, this parallels discussions in the FAIR community about challenges to FAIR implementation (such as \cite{jacobsenFAIRPrinciplesInterpretations2020}). 
	
	\subsection{FACT Principles }
	
	The four principles of responsible data science as laid out by FACT
	are Fairness, Accuracy, Confidentiality, and Transparency \cite{vanderaalstGreenDataScience2016,vanderaalstResponsibleDataScience2017}. This approach is bounded by four
	questions: Q1 (\textquotedblleft Data science without prejudice---how
	to avoid unfair conclusions even if they are true?\textquotedblright )
	maps indirectly into our Societal Values component. Q2, \textquotedblleft Data science without guesswork - how to answer questions with a guaranteed level of accuracy?\textquotedblright{} is a bit harder to map but mostly fits within Transparency. Two questions (Q3, \textquotedblleft Data science that ensures confidentiality---how to answer questions without revealing secrets?\textquotedblright{} and Q4, \textquotedblleft Data science that provides transparency---how to clarify answers so that
	they become indisputable?\textquotedblright ) can be clearly mapped
	to respectively the Privacy/Confidentiality and the Transparency component of our responsibility matrix. We loosely can map the 4 questions into the responsibility Matrix (See the red cells in Fig. \ref{fig:1} Q1-4). They are only related to the Outcomes of a project.The problem with these \textquotedblleft guiding\textquotedblright{} questions is that they do not clearly articulate how they apply to the data science project lifespan, nor is there a vision about who has responsibility in organisations for providing sufficient answers to these questions.

	\subsection{Datasheets for Datasets}
	
	Datasheets for Datasets \cite{gebruDatasheetsDatasets2021} proposes a set of questions divided into 7 categories (1.Motivation; 2. Composition; 3. Collection Process; 4. Preprocessing/Cleaning/Labelling; 5. Uses; 6. Distribution and 7. Maintenance (see Appendix \ref{appendixC}). Each category has several questions to be answered in order to complete the datasheet. 
	
	\begin{figure}
	\centering
	\includegraphics[width=0.6\textwidth]{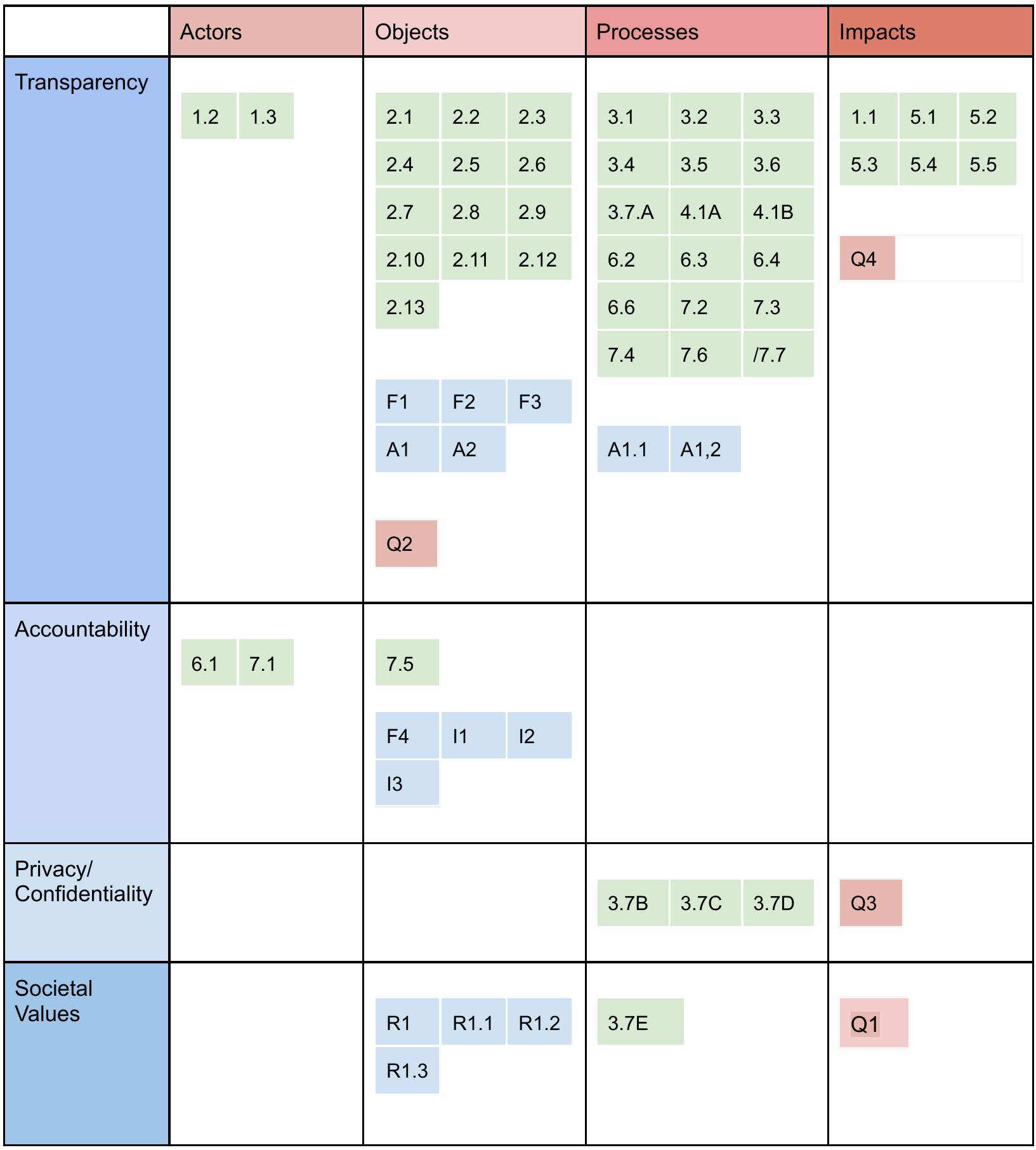}
	\caption{The resulting mapping into the TAPS responsibility matrix.}
	\label{fig:1}
    \end{figure}
	
	The green squares in \textbf{ Fig. \ref{fig:1} } show the mapping of the Datasheets for Datasets within the responsibility Matrix. Being focused on datasets only, we can conclude that the tool covers very well the transparency dimension of the Matrix. We also observe that some dimensions are not covered at all. Dimensions such as privacy and confidentiality of actors may understandably not be highly relevant in the context of data reuse, assuming that the datasets remove identifying information. Other dimensions however such as the social impact of the defined objects remain unexplored. Considering that datasheets are specifically thought of as a way to share and re-use datasets (here identified as objects), identifying the values governing data re-use could be significant. 
	
	
	\subsection{Summary of the Results }
	
The results of these mappings should not suggest that the FAIR, FACT, and Datasheets for Datasets are insufficient for the purposes for which they were designed. They do, however, suggest that even data scientists working on responsibility have some blind spots or other potential parts of the responsibility agendas that they do not pay attention to. We envisage several possible use cases where the matrix can become valuable: 
	
\begin{enumerate}
\item As a tool for planning projects. Addressing the issues in each cell and answering the question prompts can be helpful for planning a project, pulling together resources, and plotting its research agenda. Indeed, answering the questions posed in each cell of the matrix is an important part of the \textquotedblleft by design.\textquotedblright{} We note that although some of the cells may appear redundant because they pose questions about similar matters, the answers may be different depending on 1) the specific intersection of components and scopes and 2) the relationships with other cells.

\item As an audit or assessment of planned or existing projects. The matrix might be used as the basis for a data ethics review, similar to a research ethics review. 

\item As analogous to a \textquotedblleft risk register.\textquotedblright{} A risk register is a standard tool in project management for tracking ongoing and emerging risk factors and events in a project. Similarly, a \textquotedblleft responsibility register\textquotedblright{} might be a living document of the evolution of a project\textquoteright s responsibility agenda over time. Additionally, the matrix may not be filled by only one person, but rather requires a team\textquoteright s input. 
\end{enumerate}

\section{Discussion}
\label{sec:discussion}
	
Building responsibility matrices for data science projects is a way to build a responsibility agenda into data science work. This is important for a number of reasons: Moral intuitions have not developed in a vacuum. As a historically interdisciplinary field, data science can draw on its interactions with many different fields and their approaches to responsibility. This also gives data science a unique perspective on the way different communities behave towards data and thus the need to acknowledge community needs and values in the data science life cycle. Additionally, executing a  responsibility agenda in data science will require interdisciplinary collaborations with ethicists, sociologists, and legal scholars, for which data scientists are uniquely suited. 
	
\noindent A responsibility agenda rooted in data science projects can bridge and ameliorate conflicts between communities and domains which have different and divergent values around the main four elements of a responsibility agenda: transparency, privacy, accountability, and impact on society. Because private companies and governments may have different ideas about responsibility than academia and health care, necessary collaborations can be limited. Take the case of health data collected by wearable medical sensors and incorporated into electronic health records (these records are accessible by individuals, but sensor companies, not wearers, \textquotedblleft own\textquotedblright {} these data). 
Data science\textquoteright s ideas about responsibility can represent a stable common ground for future collaborations. As data science programs aimed at producing data scientists multiply around the world, it is necessary to think about a range of attributes and practices that data science possesses, which distinguishes itself from other fields. We believe it is time for data science to begin developing its own approaches to responsibility, and in particular using its expertise with data life cycles to promote responsibility within society. To put it another way, data science can be a bucket, but that does not mean the bucket is without moral values.
	
	\noindent The history of society and technology teaches that technological innovation occurs within social contexts whose values and legal frameworks change over time, sometimes rapidly and sometimes in subtle ways. It also teaches that innovation practitioners do so unaware of that context, especially when it shifts, which makes them unreliable sources of insight into ethical boundaries. The matrix allows for specific and contextual questions to be asked about a given responsibility agenda without requiring that the agenda take any particular shape. 
	\noindent In this work, we evaluated other frameworks with the purpose of mapping other guidelines onto the matrix. The mapping identifies some of the shortcomings of the ethical statements and guidelines of these frameworks, especially around the clarity and coverage of the different components. Ultimately, the matrix helps to identify the gaps and how to make better suited guidelines for the real world contexts. 
	\noindent The presented matrix also has some limitations, especially for end-users. We tested the components of the responsibility matrix with different end-users, from different domains, and they found it difficult to fill it in when no questions were presented. The matrix was then extended to include several questions, as presented in Appendix A, for each of the dimensions. Some of the questions may still be difficult to interpret and complete. In future work we are addressing these limitations by working with focus groups and designing a web-based platform for filling in the
	matrix. 
	
	\section{Conclusion and future work}
	\label{sec:conclusion}

In the data science domain, responsible design has become the next challenge \cite{stoyanovichFidesPlatformResponsible2017a}. We analyzed data science concepts artifacts in terms of scopes and components of responsibility. The result is a two-dimensional matrix of spaces that we call TAPS \textquotedblleft responsibility matrix\textquotedblright. It clarifies various questions about responsibility and uncovers issues which often remain implicit in other open science frameworks. The matrix provides an overview of how the different responsibilities are handled throughout a project life cycle for all the involved actors. Using the proposed Responsibility Matrix we evaluated existing frameworks such as FAIR, FACT and Datasheets for Datasets. We showed how the different questions raised by these frameworks can be supported within the Responsibility Matrix to best frame the responsibility dimensions of a project. Our analysis shows that the different frameworks cover different aspects of responsibility but not all. We conclude that the Responsibility Matrix can assist data scientists to evaluate data science activities in the responsibility context. 
As a future work, we are developing a web-based evaluation tool to enable data scientists to use the responsibility matrix. The web-based tool will be evaluated interactively with study groups and later enrolled to be tested by a broad audience. This will enable us to evaluate the usefulness of the TAPS model for a broad range of projects and identify further improvements to the model. Furthermore, we are developing a dictionary for supporting a shared set of concepts and relationships in completing the matrix as well as supporting the exploration of the definitions of responsibility in practice. Finally, the responsibility matrix tool can be applied to a heterogeneous range of projects. The evaluations from data scientists working on diverse domains are necessary for testing the responsibility agenda in various contexts.

	
	
	
\section*{Acknowledgement(s)}
This work is dedicated to Dr. Amrapali Zaveri. Thank you to Prof. David Townend and Dr. Birgit Wouters for their advice into the privacy and confidentiality dimensions of the matrix. This work was partially funded by the Maastricht University Ethics and Integrity Grant, York-Maastricht Partnership, the Digital Society Research Programme, and the NWO-Aspasia Grant (no. 91716421).

\bibliographystyle{plain}
\bibliography{RDS2022Paper}


	\appendix

	\section{TAPS Matrix - Questions}
	Appendix A : TAPS Matrix - Questions
    \label{appendixA}
    

| Question | ID | Component | Scope |
|---|---|---|---|
| Who is involved in this project? e.g. researchers, organizations, end users | TA1 | Transparency | Actors |
| Have all the humans involved in the project been identified? This includes data scientists, downstream users of data science objects, & data subjects (e.g., patients). | TA2 | Transparency | Actors |
| Are there formal agreements or explicitly described relationships between them? Examples of formal agreements: employment contracts (specifically length of contract, percentage of work, employee classifications, & other characteristics that might interact with ACCOUNTABILITY), consent forms, ethical approvals, intellectual property protocols, patents, non-disclosure agreements, contracts governing engagements | TA3 | Transparency | Actors |
| Are these formal agreements accessible by all of the actors? | TA4 | Transparency | Actors |
| Have the contributions of the humans who are involved been described? | TA5 | Transparency | Actors |
| Are there mechanisms in place to mitigate potential conflicts between actors? | TA6 | Transparency | Actors |
| Is there a provision for continuity and sustainability of the project and its resources over time? (e.g., do contact emails go to institutional email addresses, not individuals'?) | TA7 | Transparency | Actors |
| Are there any other secondary actors involved or affected by the project? (e.g., investors, funding agency) | TA8 | Transparency | Actors |
| What the outputs or deliverables of the project are? e.g. types of data to collect/re-use, models or algorithms | TO1 | Transparency | Objects |
| Have all data/metadata/algorithms been described [this may be a link to a document] | TO1 | Transparency | Objects |
| How will you make data and algorithms you used in the analysis transparent? | TO2 | Transparency | Objects |
| How will you make data and algorithms that you used in the analysis accessible to others? (such as following the FAIR principles, meeting Open Science guidelines) | TO3 | Transparency | Objects |
| Is any metadata given for the dataset? | TO4 | Transparency | Objects |
| Are your data and metadata in a repository? (provide links such as Github, Zenodo etc.) | TO5 | Transparency | Objects |
| Are the data and metadata publicly available? | TO6 | Transparency | Objects |
| What is the licence associated with the data? | TO7 | Transparency | Objects |
| Have you measured the quality of the data? If yes, what quality dimensions and metrics did you use (i.e. accuracy, completeness, consistency, timeliness, uniqueness, and validity )? What is the overall quality of the data? | TO8 | Transparency | Objects |
| What is the quantity of missing values in the data? | TO9 | Transparency | Objects |
| Are there any terms and conditions specified on the use of the data/algorithm? | TO10 | Transparency | Objects |
| Describe the overall process used to achieve the project's outputs or deliverables (including data science lifecycle and project planning)? e.g. data collection, cleaning, data management | TP1 | Transparency | Process |
| How will you collect the data? | TP2 | Transparency | Process |
| How do you clean/preprocess the data? | TP3 | Transparency | Process |
| How do you analyse the data? | TP4 | Transparency | Process |
| How do you maintain/serve the project or your data? | TP5 | Transparency | Process |
| What are the aims of the project? | TI1 | Transparency | Impact |
| What are the outcomes of the project? | TI2 | Transparency | Impact |

| Question | Code | Category | Type |
|---|---|---|---|
| What are the expected impacts of the project? | TI3 | Transparency | Impact |
| Can you describe the implications of the project so that it is understandable by everyone? | TI4 | Transparency | Impact |
| Can you articulate how this project will affect people's life or research? | TI5 | Transparency | Impact |
| Have you described how this project might be misused? | TI6 | Transparency | Impact |
| Are there any terms and conditions specified on the use of data/algorithm? | TI7 | Transparency | Impact |
| Has the researcher signed a code of conduct agreement? | AA1 | Accountability | Actors |
| Is it clear who is responsible for which part of the outcome? | AA2 | Accountability | Actors |
| Does your organisation have protocols and procedures for dealing with an intentional or unintentional breakdown of transparency, privacy/confidentiality, or norm breaking? | AA3 | Accountability | Actors |
| Are there protocols and procedures for dealing with conflicts between legal and ethical positions in projects? | AA4 | Accountability | Actors |
| Does your organisation have a data protection officer? Data ethics review committees? Data stewards? | AA5 | Accountability | Actors |
| Do employees have access to resources (e.g., training) for doing responsible data science? | AA6 | Accountability | Actors |
| Do protections exist against lawsuit or damages for individual data scientists, users, or organisations? | AA7 | Accountability | Actors |
| What is the provenance of the data? | AO1 | Accountability | Objects |
| Do you know the original permissions attached to the data (in the case of re-use)? | AO2 | Accountability | Objects |
| Are you following the original permissions that are attached to that data? | AO3 | Accountability | Objects |
| Is there a description provided about potential biases, such as confounding factors? | AO4 | Accountability | Objects |
| Is data quality an issue? If yes, what is the specific issue and how is it resolved? | AO5 | Accountability | Objects |
| Is there any missing data or missing variables and how are these values handled? | AO6 | Accountability | Objects |
| Is there a clear, transparent and accessible description of the data collection and data management processes in the project? | AP1 | Accountability | Process |
| How responsibilities for the processes (data collection, management or analysys) are assigned? | AP2 | Accountability | Process |
| Is there any control/checkpoints within the process? | AP3 | Accountability | Process |
| What is the impact or the outcome of the analysis? | AI1 | Accountability | Impact |
| How are the outcomes of the project communicated to the stakeholders? | AI2 | Accountability | Impact |
| How are unintended outcomes communicated to the relevant organisation? | AI3 | Accountability | Impact |
| Are there protocols and procedures for dealing with conflicts between legal and ethical positions in the project? | AI4 | Accountability | Impact |
| Is there a procedure for dealing with unintended outcomes, either positive or negative? | AI5 | Accountability | Impact |
| How does your organisation deal with evaluations by external parties about the responsibility aspects of the project? | AI6 | Accountability | Impact |
| Is there an informed consent form signed by the human subjects in the project? | PA1 | Privacy | Actors |
| Has a confidentiality agreement or non disclosure agreement been signed (if applicable)? | PA2 | Privacy | Actors |
| Has a data usage agreement been signed by the researchers involved? | PA3 | Privacy | Actors |

| Question | Code | Category | Type |
|---|---|---|---|
| Has a privacy protection method been used to make the data anonymous or pseudonymous? | PO1 | Privacy | Objects |
| Are you collecting identifiable personal information? | PO2 | Privacy | Objects |
| Are you collecting sensitive data (as defined by GDPR)? | PO3 | Privacy | Objects |
| Are the data you collect or re-use necessary to answer your research questions? | PO4 | Privacy | Objects |
| Where/how do you process the data? (do you upload data to some servers to execute processing? Which tools do you use?) | PO5 | Privacy | Objects |
| Where do you store data during and after your project? | PO6 | Privacy | Objects |
| Is it secure? | PO7 | Privacy | Objects |
| Did you do your data management plan or use a template (such as https://dmponline.dcc.ac.uk/)? | PO8 | Privacy | Objects |
| Can third parties use the data? if so, how? | PO9 | Privacy | Objects |
| Do your models and results reveal personal information directly? | PO10 | Privacy | Objects |
| Can the outcome be used to identify individuals indirectly? | PO11 | Privacy | Objects |
| Is it possible that other people can combine your model/algorithm with other resources to identify individuals? | PO12 | Privacy | Objects |
| Did you collect the data from the individuals in question directly, or obtain it via third parties or other sources (e.g., websites)? | PP1 | Privacy | Process |
| Were the individuals in question notified about the data collection? If so, please describe (or show with screenshots or other information) how notice was provided, and provide a link or other access point to, or otherwise reproduce, the exact language of the notification itself | PP2 | Privacy | Process |
| Did the individuals in question consent to the collection and use of their data? If so, please describe how consent was requested and provided, and provide a link or other access point to, or otherwise reproduce, the exact language to which the individuals consented. | PP3 | Privacy | Process |
| If consent was obtained, were the consenting individuals provided with a mechanism to revoke their consent in the future or for certain uses? If so, please provide a description, as well as a link or other access point to the mechanism (if appropriate). | PP4 | Privacy | Process |
| What procedures were used to ensure the anonymity of personal data? For example, removing sensitive data fields, aggregating at a group level, data encryption mechanisms…ect | PP5 | Privacy | Process |
| Has an information management system been set in place for auditing purposes? | PP6 | Privacy | Process |
| Has an analysis of the potential impact of the dataset and its use on data subjects (e.g., a data protection impact analysis) been conducted? If so, please provide a description of this analysis, including the outcomes, as well as a link or other access point to any supporting documentation | PI1 | Privacy | Impact |
| Does your project have the potential to impact the privacy and confidentiality of individuals? If yes, how? | PI2 | Privacy | Impact |
| Can your project change society's confidence and trust in data science projects? If yes, how? | PI3 | Privacy | Impact |
| Will your outcome deliver a positive message to people about how their privacy and confidentiality have been handled? | PI4 | Privacy | Impact |
| What are the operative social values and norms of the actors that are involved in the project? (i.e. cultural or ethical norms) | SA1 | Societal Values | Actors |

| Question | ID | Category | Subcategory |
|---|---|---|---|
| In which country(is) or location(s) is the work being conducted (Europe, US, Asia, etc)? | SA2 | Societal Values | Actors |
| Are there any code of conducts that your project should follow? | SA3 | Societal Values | Actors |
| What are the operative social values and norms that touch on data, algorithms, and other outputs or deliverables of the project? | SO1 | Societal Values | Objects |
| Are there any laws or regulations in place to secure data protection (like the GDPR, CCPA in California)? | SO2 | Societal Values | Objects |
| What are the practices for sharing research data with the community that your project strives to follow? | SO3 | Societal Values | Objects |
| Is there a data sharing agreement between the partners within or beyond the project? | SO4 | Societal Values | Objects |
| Are there any governing rules affecting your project in regards to use of devices and/or digital applications? | SO5 | Societal Values | Objects |
| Are there any rules or standards for data collection? | SP1 | Societal Values | Process |
| Did you pre-register your study somewhere (share your research data management plan)? For example via Center for Open Science (OSF)[1]. | SP3 | Societal Values | Process |
| Does the data analysis process affect specific population groups? | SP4 | Societal Values | Process |
| If you are collecting data, what are the steps taken to reduce possible societal bias (i.e gender, population, cultural, age biases) | SP5 | Societal Values | Process |
| If you are using already collected data, what are the steps taken to reduce possible societal biases existing in these data (i.e gender, population, cultural, age biases)? | SP6 | Societal Values | Process |
| What are the societal impacts of this project? | SI1 | Societal Values | Impact |
| Will the project inform the stakeholders about any possible side-effects or negative impacts? | SI2 | Societal Values | Impact |
| What could be possible commercial interests in the project's outcomes? | SI3 | Societal Values | Impact |
| What is the public interest in the project's outcomes once they mature? | SI4 | Societal Values | Impact |
| Is there an identifiable conflict between societal values of the commercial and public interests in the outcomes? | SI5 | Societal Values | Impact |
| Are there any procedures or norms in place to resolve different interests between public interest and commercial interests? | SI6 | Societal Values | Impact |
| Did you create a Data protection impact assessment (DPIA)[2] which is obligatory under the GDPR in the Netherlands. | SI7 | Societal Values | Impact |
| Are all of the operative social values and norms in agreement with each other? If not please explain. | SI8 | Societal Values | Impact |
| Can you describe or estimate the environmental impact of the project (for example: energy consumption)? | SI9 | Societal Values | Impact |
| Does the project rely on the use of any toxic or non recyclable materials? | SI10 | Societal Values | Impact |

---

[1] Link: https://osf.io/x5w7h/wiki/06%20LeaderBoard/?hsCtaTracking=2f25be73-35de-4b95-93e8-4f08acc42332%7Cf2024186-54bb-4626-baab-1785d2f9d71a

[2] Link: https://autoriteitpersoonsgegevens.nl/sites/default/files/atoms/files/stcrt-2019-64418.pdf


    \section{The FAIR Guiding Sub-principles}
    \label{appendixB}
    
    Appendix B
The FAIR Guiding Principles

Findability:
\begin{itemize}
    \item F1. (meta)data are assigned a globally unique and persistent identifier
   \item  F2. data are described with rich metadata (defined by R1 below)
   \item  F3. metadata clearly and explicitly include the identifier of the data it describes
    \item F4. (meta)data are registered or indexed in a searchable resource
\end{itemize}

Accessibility:
\begin{itemize}
\item A1. (meta)data are retrievable by their identifier using a standardized communications
protocol
\item A1.1 the protocol is open, free, and universally implementable
\item A1.2 the protocol allows for an authentication and authorization procedure, where necessary
\item A2. metadata are accessible, even when the data are no longer available
\end{itemize}

Interoperability:
\begin{itemize}
\item I1. (meta)data use a formal, accessible, shared, and broadly applicable language for
knowledge representation.
\item I2. (meta)data use vocabularies that follow FAIR principles
\item I3. (meta)data include qualified references to other (meta)data
\end{itemize}

Reusability:
\begin{itemize}
\item R1. meta(data) are richly described with a plurality of accurate and relevant attributes
\item R1.1. (meta)data are released with a clear and accessible data usage license
\item R1.2. (meta)data are associated with detailed provenance
\item R1.3. (meta)data meet domain-relevant community standard
\end{itemize}

\section{Datasheets for datasets - Questions}
\label{appendixC}
The following is an extract of the questions presented by Gebru et al in the data sheets for datasets paper \cite{gebruDatasheetsDatasets2021}. We show how we labeled each question in order to map them into the matrix and the questions as stated in Gebru et al work. Some example statements have been
shortened. The questions are numbered in the order of appearance within the categories. For example, the first question of the motivation category is labelled 1.1 while the last question in the maintenance category is labelled 7.8 (this category has 8 questions). If questions were accompanied by subsections, they were labelled with letters (i.e 4.1, 4.1.A and 4.1.B). If the subsections were placed on the same matrix cell, then only the main label of the question is used (i.e is 4.1 and all the sub questions belong to the cell transparency/process, then only 4,1 is noted in the mapping, however the sub-questions of 3.7
could not be placed in the same cell as they relate to different scopes, in the matrix, then we noted the
exact placement of each sub-question within the matrix). Every section ended with: “Any other
comments?”, these questions were labeled but not deemed relevant to be mapped within the matrix.
\begin{enumerate}
    \item \textbf{Motivation}
        \begin{itemize}
            \item \textbf{1.1} For what purpose was the dataset created? Was there a specific task in mind? Was there a specific gap that needed to be filled? Please provide a description.
            \item \textbf{1.2} Who created the dataset (e.g., which team) and on behalf of which entity (e.g., institution)?
            \item \textbf{1.3} Who funded the creation of the dataset? If there is an associated grant, please provide the name of the grantor and the grant name and number.
            \item 1.4 Any other comments? (Not mapped due to being a broad question)
        \end{itemize}
    \item \textbf{Composition}
    \begin{itemize}
        \item \textbf{2.1} What do the instances that comprise the dataset represent (e.g., documents, photos, people, countries)? Are there multiple types of instances (e.g., movies, users, and ratings)? Please provide a description.
        \item \textbf{2.2} How many instances are there in total (of each type, if appropriate)?
        \item \textbf{2.3} Does the dataset contain all possible instances or is it a sample (not necessarily random) of instances from a larger set? If the dataset is a sample, then what is the larger set? Is the sample representative of the larger set (e.g., geographic coverage)? If so, please describe how this representativeness was validated/verified. If it is not representative of the larger set, please describe why not (e.g., to cover a more diverse range of instances, because instances were withheld).
        \item \textbf{2.4} What data does each instance consist of? “Raw” data (e.g., unprocessed text or images) or features? In either case, please provide a description.
        \item \textbf{2.5} Is there a label or target associated with each instance? If so, please provide a description.
        \item \textbf{2.6} Is any information missing from individual instances? If so, please provide a description, explaining why this information is missing (e.g., because it was unavailable). This does not include intentionally removed information, but might include, e.g., redacted text.
        \item \textbf{2.7} Are relationships between individual instances made explicit (e.g., users’ movie ratings, social network links)? If so, please describe how these relationships are made explicit.
        \item \textbf{2.8} Are there recommended data splits (e.g., training, development/validation, testing)? If so, please provide a description of these splits, explaining the rationale behind them.
        \item \textbf{2.9} Are there any errors, sources of noise, or redundancies in the dataset? If so, please provide a description.
        \item \textbf{2.10} Is the dataset self-contained, or does it link to or otherwise rely on external resources (e.g., websites, tweets, other datasets)? If it links to or relies on external resources, a) are there guarantees that they will exist, and remain constant, over time; b) are there official archival versions of the complete dataset (i.e., including the external resources as they existed at the time the dataset was created); c) are there any restrictions (e.g., licenses, fees) associated with any of the external resources that might apply to a future user? Please provide descriptions of all external resources and any restrictions associated with them, as well as links or other access points, as appropriate.
        \item \textbf{2.11} Does the dataset contain data that might be considered confidential (e.g., data that is protected by legal privilege or by doctor patient confidentiality, data that includes the content of individuals’ non-public communications)? If so, please provide a description.
        \item \textbf{2.12} Does the dataset contain data that, if viewed directly, might be offensive, insulting, threatening, or might otherwise cause anxiety? If so, please describe why.
        \item 2.13 Does the dataset relate to people? If not, you may skip the remaining questions in this section.
        \item \textbf{2.13A} Does the dataset identify any subpopulations (e.g., by age, gender)? If so, please describe how these subpopulations are identified and provide a description of their respective distributions within the dataset.
        \item \textbf{2.13B} Is it possible to identify individuals (i.e., one or more natural persons), either directly or indirectly (i.e., in combination with other data) from the dataset? If so, please describe how.
        \item \textbf{2.13C} Does the dataset contain data that might be considered sensitive in any way (e.g., data that reveals racial or ethnic origins, sexual orientations, religious beliefs, political opinions or union memberships, or locations; financial or health data; biometric or genetic data; forms of government identification, such as social security numbers; criminal history)? If so, please provide a description.
        \item 2.14 Any other comments? (Not mapped due to being a broad question)
    \end{itemize}
    \item \textbf{Collection Process}
    \begin{itemize}
        \item \textbf{3.1} How was the data associated with each instance acquired? Was the data directly observable (e.g., movie ratings), reported by subjects (e.g., survey responses), or indirectly inferred/derived from other data (e.g., part-of-speech tags)? If data was reported by subjects or indirectly inferred/derived from other data, was the data validated/verified? If so, please describe how.
        \item \textbf{3.2} What mechanisms or procedures were used to collect the data (e.g., sensor, manual human curation, software program)? How were these mechanisms or procedures validated?
        \item \textbf{3.3} If the dataset is a sample from a larger set, what was the sampling strategy (e.g., deterministic, probabilistic with specific sampling probabilities)?
        \item \textbf{3.4} Who was involved in the data collection process (e.g., students, crowd-workers) and how were they compensated (e.g., how much were crowd-workers paid)?
        \item \textbf{3.5} Over what timeframe was the data collected? Does this timeframe match the creation timeframe of the data associated with the instances (e.g., recent crawl of old news articles)? If not, please describe the timeframe in which the data associated with the instances was created.
        \item \textbf{3.6} Were any ethical review processes conducted (e.g., by an institutional review board)? If so, please provide a description of these review processes, including the outcomes, as well as a link or other access point to any supporting documentation.
        \item 3.7 Does the dataset relate to people? If not, you may skip the remainder of the questions in this section.
        \item \textbf{3.7A} Did you collect the data from the individuals in question directly, or obtain it via third parties or other sources (e.g., websites)?
        \item \textbf{3.7B} Were the individuals in question notified about the data collection? If so, please describe (or show with screenshots or other information) how notice was provided, and provide a link or other access point to, or otherwise reproduce, the exact language of the notification itself.
        \item \textbf{3.7C} Did the individuals in question consent to the collection and use of their data? If so, please describe (or show with screenshots or other information) how consent was requested and provided, and provide a link or other access point to, or otherwise reproduce, the exact language to which the individuals consented.
        \item \textbf{3.7D} If consent was obtained, were the consenting individuals provided with a mechanism to revoke their consent in the future or for certain uses? If so, please provide a description, as well as a link or other access point to the mechanism (if appropriate).
        \item \textbf{3.7E} Has an analysis of the potential impact of the dataset and its use on data subjects (e.g., a data protection impact analysis)been conducted? If so, please provide a description of this analysis, including the outcomes, as well as a link or other access point to any supporting documentation.
        \item 3.8 Any other comments? (Not mapped due to being a broad question)
    \end{itemize}
    \item \textbf{Preprocessing/Cleaning/Labeling}
    \begin{itemize}
        \item 4.1 Was any preprocessing/cleaning/labeling of the data done (e.g., discretization or bucketing, tokenization, removal of instances, processing of missing values)? If so, please provide a description.
        \item \textbf{4.1A} Was the “raw” data saved in addition to the preprocessed/cleaned/labeled data (e.g., to support unanticipated uses)? If so, please provide a link or other access point to the “raw” data. 
        \item \textbf{4.1B} Is the software used to preprocess/clean/label the instances available? If so, please provide a link or other access point.
        \item 4.2 Any other comments? (Not mapped due to being a broad question)
    \end{itemize}
    \item \textbf{Uses}
    \begin{itemize}
        \item \textbf{5.1} Has the dataset been used for any tasks already? If so, please provide a description.
        \item \textbf{5.2} Is there a repository that links to any or all papers or systems that use the dataset? If so, please provide a link or other access point.
        \item \textbf{5.3} What (other) tasks could the dataset be used for?
        \item \textbf{5.4} Is there anything about the composition of the dataset or the way it was collected and preprocessed/cleaned/labeled that might impact future uses? For example, is there anything that a future user might need to know to avoid uses that could result in unfair treatment of individuals or groups (e.g., stereotyping, quality of service issues) or other undesirable harms (e.g., financial harms, legal risks) If so, please provide a description. Is there anything a future user could do to mitigate these undesirable harms?
        \item \textbf{5.5} Are there tasks for which the dataset should not be used? If so, please provide a description.
        \item 5.6 Any other comments? (Not mapped due to being a broad question)
    \end{itemize}
    \item \textbf{Distribution}
    \begin{itemize}
        \item \textbf{6.1} Will the dataset be distributed to third parties outside of the entity (e.g., company, institution, organization) on behalf of which the dataset was created? If so, please provide a description.
        \item \textbf{6.2} How will the dataset will be distributed (e.g.,website, API, GitHub)? Does the dataset have a digital object identifier (DOI)?
        \item \textbf{6.3} When will the dataset be distributed?
        \item \textbf{6.4} Will the dataset be distributed under a copyright or other intellectual property (IP) license, and/or under applicable terms of use (ToU)? If so, please describe this license and/or ToU, and provide a link or other access point to, or otherwise reproduce, any relevant licensing terms or ToU, as well as any fees associated with these restrictions.
        \item \textbf{6.5} Have any third parties imposed IP-based or other restrictions on the data associated with the instances? If so, please describe these restrictions, and provide a link or other access point to, or otherwise reproduce, any relevant licensing terms, as well as any fees associated with these restrictions.
        \item \textbf{6.6} Do any export controls or other regulatory restrictions apply to the dataset or to individual instances? If so, please describe these restrictions, and provide a link or other access point to, or otherwise reproduce, any supporting documentation.
        \item 6.7 Any other comments? (Not mapped due to being a broad question)
    \end{itemize}
    \item \textbf{Maintenance}
    \begin{itemize}
        \item \textbf{7.1} Who is supporting/hosting/maintaining the dataset?
        \item \textbf{7.2} How can the owner/curator/manager of the dataset be contacted (e.g., email address)?
        \item \textbf{7.3} Is there an erratum? If so, please provide a link or other access point.
        \item \textbf{7.4} Will the dataset be updated (e.g., to correct labeling errors, add new instances, delete instances)? If so, please describe how often, by whom, and how updates will be communicated to users (e.g., mailing list, GitHub)? 
        \item \textbf{7.5} If the dataset relates to people, are there applicable limits on the retention of the data associated with the instances (e.g., were individuals in question told that their data would be retained for a fixed period of time and then deleted)? If so, please describe these limits and explain how they will be enforced.
        \item \textbf{7.6} Will older versions of the dataset continue to be supported/hosted/maintained? If so, please describe how. If not, please describe how its obsolescence will be communicated to users. \item \textbf{7.7} If others want to extend/augment/build on/contribute to the dataset, is there a mechanism for them to do so? If so, please provide a description. Will these contributions be validated/verified? If so, please describe how. If not, why not? Is there a process for communicating/distributing these contributions to other users? If so, please provide a description.
        \item 7.8 Any other comments? (Not mapped due to being a broad question) 
    \end{itemize}
\end{enumerate}

    
\end{document}